# Influence of atomic collisions on spectrum of light scattered from an $f$-deformed Bose-Einstein condensate

Z. Haghshenasfard[1], M. H. Naderi[2] and M. Soltanolkotabi[3]
*Quantum Optics Group, Department of Physics, University of Isfahan,
Hezar Jerib, Isfahan, 81746-73441, Iran*

**Abstract**
In this paper, we investigate the spectrum of light scattered from a Bose-Einstein condensate in the framework of $f$-deformed boson. We use an $f$-deformed quantum model in which the Gardiner's phonon operators for BEC are deformed by an operator-valued function, $f(\hat{n})$, of the particle-number operator $\hat{n}$. We consider the collisions between the atoms as a special kind of $f$-deformation. The collision rate $\kappa$ is regarded as the deformation parameter and the spectrum of light scattered from the deformed BEC is analyzed. In particular, we find that with increasing the values of deformation parameters $\kappa$ and $\eta = \frac{1}{N}$ ($N$, total number of condensate atoms) the scattering spectrum shows deviation from the spectrum associated with nondeformed Bose-Einstein condensate.

# 1 Introduction

The recent experimental realization of Bose-Einstein condensate (BEC) [1-5] has generated much interest in studying the properties of a BEC. One question that has received much attention is what signatures of BEC imprint in the spectrum of light scattered from BEC [6-13]. Recently, Javanainen [6] has been calculated the spectrum of light scattered from a low density degenerate Bose gas, in the limit of large detuning. He showed that the spectrum may display additional qualitative features arising directly from Bose-Einstein statistics and identified a two-peak structure of the scattering function, one at frequencies below the incident light frequency due to the recoil of the gas atoms by the change in the wavevector of the photon in the scattering process and the other above the incident light frequency arising from the preferred scattering of an atom to an already occupied state of the BEC. Light scattering from a weakly interacting BEC has been considered by Graham and Walls [7]. They found that at large detuning, light scattering shows two symmetrically placed peaks whose intensities are related by detailed balance and their frequencies depend on the product of the scattering length and total number of atoms within the condensate. The weak coherent light scattering off a BEC in the limit of a very large trap ($\frac{a}{\lambdabar} \gg 1, \frac{N\lambdabar^3}{a^3} \gg 1$ where $N$ denotes total number of condensate atoms, $\lambdabar$ is the resonance wavelength and $a$ is the size of the trap's ground state) have been considered in [8, 9]. It has been found that there exists a gap in the excitation spectrum, due to mixing of the atomic and photonic degrees of freedom. Because of this gap the resonant light will be strongly reflected back from the sharp boundary of the condensate. Induced and spontaneous optical processes in a condensate of an ideal (noninteracting) Bose gas in the limit of a small trap ($\frac{a}{\lambdabar} \ll 1, \frac{N\lambdabar^3}{a^3} \gg 1$) have been discussed by Javanainen [10]. He found that, in the steady state regime, a small number of atoms remain in the excited state and the number of scattered photons has a Lorentzian line shape centered at the bare atomic resonance transition frequency. You and coworkers [11] have been demonstrated that in the intermediate regime ($\frac{a}{\lambdabar} \approx 1, \frac{N\lambdabar^3}{a^3} \approx 1$) the results are significantly modified. They found that the line shape is non-Lorentzian and in some circumstances exhibits additional interesting features. Band [12] has been considered the excitation spectrum of the BEC in a harmonic oscillator potential and the external magnetic field. He found that the intensities of the scattering peaks depend strongly on temperature. The temperature dependence of the relative and absolute intensities of the peaks is a strong indication of BEC. In Ref. [13], it has been pointed out that the limiting of small number of atoms within the BEC could lead to significant changes in the spectrum of light scattered from a BEC.

---

[1] E-mail: zhaghshenas@hotmail.com
[2] E-mail: mhnaderi2001@yahoo.com
[3] E-mail: soltan@sci.ui.ac.ir

The fundamental approach in the description of BEC is the Bogoliubov method [14], in which the operators for condensate atoms are replaced by $c$-numbers. This approximation breaks symmetry of the resulting approximate Hamiltonian, i.e., the total number of atoms $N$, may not be conserved. Gardiner [15] solved this problem by introducing the exciton operators which conserve the total number of atoms $N$ and satisfy an $f$-deformed commutation relation. As $N \to \infty$ the standard bosonic commutation relation is regained. Recently, the case of the finite number of atoms has been investigated [16]. It provides a physical and natural realization of the $f$-deformed boson [16] by using the Gardiner's phonon operators for the description of BEC. In the last few years, there have attracted a great deal of interest in quantum groups and their associated algebra, which are specific deformations of Lie algebra. Algebraic models have been used very successfully in several research areas of physics and mathematics such as exactly solvable statistical models [17], non commutative geometry [18], nuclear quantum many body problems [19] and rational conformal field theories [20]. Recent interest in quantum groups has led to introduce the concept of $q$-deformed boson oscillators [21, 22, 23] which are deformations of standard bosonic harmonic oscillator algebra. The $q$-deformed boson oscillator has been interpreted as a nonlinear oscillator with a very specific type of nonlinearity which classically corresponds to an intensity dependence of the oscillator frequency [24]. In addition, a general type of nonlinearity for which the intensity dependence of frequency of oscillations is described by a generic function $f(\hat{n})$, the so-called $f$-deformation, has been introduced [25]. One of the most important properties of $f$-deformed bosons is their relation to nonlinearity of a special type [26]. The relation between deformed radiation field and nonlinear quantum optical processes has been studied [27]. Recently, new insight into description of BEC has been obtained by understanding and applying the $f$-deformed bosons. The type of quantum nonlinearity introduced by $f$-deformation provides a compact description of physical effects in BEC. For example, it has been revealed how $f$-deformed BEC produces a correction to the Plank distribution formula [24, 26]. In Ref. [26] it has been shown that the quantum nonlinearity introduced by $f$-deformation, changes the specific heat behavior. It is reasonable to expect that the spectrum of light scattered from a BEC is modified in the presence of deformations. In Ref. [13] it has been pointed out that intrinsic deformation due to the number of condensate atoms, extensively changes the spectrum of light scattered from a BEC. We have been recently shown [28] that the collisions effect transforms the standard harmonic oscillator model into an $f$-deformed one. It is useful to ask how the results of Ref. [13] are modified by considering the effect of collisions between the atoms within the condensate as a special kind of $f$-deformation.

In this paper, we are intended to study the spectrum of light scattered from atoms within an $f$-deformed BEC. The system under consideration is an $f$-deformed BEC of a trapped atomic gas composed of two-level atoms, in which the Gardiner's phonon operators for BEC are deformed by an operator-valued function $f(\hat{n})$. By considering the effect of collisions between the atoms within the condensate as a special kind of $f$-deformation, in which the collision rate $\kappa$ is regarded as the deformation parameter, we study the spectrum of light scattered from $f$-deformed BEC. Such a system offers extra degrees of flexibility $(\kappa, N)$ for the response of $f$-deformed BEC to the laser light. We show that the $f$-deformed BEC exhibits nonlinear characteristics, such that the nonlinearity increases by adjusting the deformation parameters $(\kappa, N)$ and the nonlinearity may lead to a deviation from the typical predicted spectrum's shape.

The present paper is organized as follows. In section 2 we present our model and we give an analytical expression for the light scattered from the $f$-deformed BEC. We use the deformation algebra to study the condensate with large but finite number of atoms. Here the deformation parameter is no longer phenomenological and is defined by the total number of atoms. We show that the atomic collisions within the condensate can be regarded as an extra deformation on the intrinsically deformed Gardiner's phonon operators for BEC. In the presence of atomic collisions within the condensate, we analyze the light scattering from the $f$-deformed BEC. Summary and conclusions are given in section 3

## 2 The model and analytical solution for spectrum of light scattered from an $f$-deformed BEC

We consider a system consisting of weakly interacting BEC of two-level atoms in a trap interacting resonantly with a classical radiation field. We introduce the creation and annihilation operators $\hat{b}^+(\hat{a}^+)$ and $\hat{b}(\hat{a})$, respectively, for the atoms in the excited (ground) state. The frequency of the laser field is denoted by $\bar{\omega}$. Moreover, we consider a reservoir, beside the classical driving field. The reservoir may be taken as any large collection of systems with many degrees of freedom. We assume that the reservoir consists of many phonon modes with closely spaced frequencies $\Omega_k$ and creation (annihilation) operators $\hat{c}_k^+(\hat{c}_k)$. Therefore, under the rotating wave approximation, the total Hamiltonian can be written as

$$\hat{H} = \hbar\bar{\omega}\hat{b}^+\hat{b} + \hbar[g(t)\hat{b}^+\hat{a} + g^*(t)\hat{a}^+\hat{b}] + \sum_k \Omega_k \hat{c}_k^+\hat{c}_k + \hbar\sum_k \xi(k)[\hat{b}^+\hat{a}\hat{c}_k + \hat{c}_k^+\hat{a}^+\hat{b}], \qquad (1)$$

where, $\xi(k)$ is the coupling coefficient pertaining to the internal atomic states and $g(t)$ is given by $g\exp(-i\Omega t)$, where $g$ is the coupling coefficient for the classical field.

It should be noted that in the above Hamiltonian the total number of atoms $N = \hat{b}^+\hat{b} + \hat{a}^+\hat{a}$ is conserved ($[\hat{N}, \hat{H}] = 0$). To deal with the dynamics of BEC the Bogoliubov approximation [14] is usually applied, in which the creation and annihilation operators $\hat{a}^+$ and $\hat{a}$ of the ground sate are substituted by $\sqrt{N_c}$, where $N_c$ is the number of initial condensate atoms. In this limit, the Hamiltonian (1) can be written as

$$\hat{H} = \hbar\bar{\omega}\hat{b}^+\hat{b} + \hbar\sqrt{N_c}[g(t)\hat{b}^+ + g^*(t)\hat{b}] + \sum_k \Omega_k \hat{c}_k^+\hat{c}_k + \hbar\sqrt{N_c}\sum_k \xi(k)[\hat{b}^+\hat{c}_k + \hat{c}_k^+\hat{b}]. \qquad (2)$$

The fast frequency dependence of $\hat{b}(t)$ can be eliminated by transforming to the slowly varying annihilation operator

$$\tilde{\hat{b}}(t) = \hat{b}(t)e^{i\Omega t}. \qquad (3)$$

The creation and annihilation operators $\tilde{\hat{b}}^+(t)$ and $\tilde{\hat{b}}(t)$ obey the bosonic commutation relation

$$[\tilde{\hat{b}}^+(t), \tilde{\hat{b}}(t)] = 1 \qquad (4)$$

We can therefore write the Heisenberg equation of motion for the operator $\tilde{\hat{b}}(t)$ in the following form

$$\dot{\tilde{\hat{b}}}(t) = -i\Delta\tilde{\hat{b}}(t) - ig\sqrt{N_c} + \sqrt{2\Gamma}\hat{b}_{in}(t) - \Gamma\tilde{\hat{b}}(t), \qquad (5)$$

where $\Delta = \bar{\omega} - \Omega$, the damping rate $\Gamma$ is given by $\Gamma = \gamma\sqrt{N_c}$ [6] with $\gamma$ as the one-atom line width and $\hat{b}_{in}(t)$ is the vacuum noise operator, having the following correlation properties

$$\left\langle \hat{b}_{in}^+(t)\hat{b}_{in}^+(t')\right\rangle = \left\langle \hat{b}_{in}(t)\hat{b}_{in}(t')\right\rangle = 0,$$
$$\left\langle \hat{b}_{in}(t)\hat{b}_{in}^+(t')\right\rangle = \delta(t-t'). \qquad (6)$$

In steady-state regime, by using semiclassical approximation $\tilde{\hat{b}}(t) = \beta_0 + \delta\tilde{\hat{b}}(t)$, Eq. (5) can be solved and we get

$$\left\langle \tilde{\hat{b}}(t) \right\rangle \equiv \beta_0 = \frac{-ig\sqrt{N_c}}{\Gamma + i\Delta}, \tag{7}$$

$$\delta\tilde{\hat{b}}(t) = \frac{\sqrt{2\Gamma}}{\Gamma + i\Delta}\hat{b}_{in}(t). \tag{8}$$

Therefore we obtain the Fourier component of the operator $\delta\tilde{\hat{b}}(t)$ in the following form

$$\delta\tilde{\hat{b}}(\omega) = \frac{\sqrt{2\Gamma}}{\Gamma + i\Delta}\hat{b}_{in}(\omega). \tag{9}$$

Now, we give an analytical expression for the spectrum of light scattered from the BEC. According to [6], the spectrum has been obtained by calculating correlation functions for the matter field operators. Hence in the steady-state regime, the standard definition of the spectrum is

$$S(\omega) = \int d\omega' \left\langle \delta\tilde{\hat{b}}^+(\omega)\delta\tilde{\hat{b}}(\omega') \right\rangle. \tag{10}$$

By using Eqs. (6) and (10) we obtain $S(\omega) = 0$. This means that in the long time limit, only the equal time correlations survive.

As pointed out before, to deal with the dynamics of BEC the Bogoliubov approximation is usually applied. One consequence of this approximation is that the total number of atoms may not be conserved. To avoid this difficultly, the following Gardiner's phonon operators are introduced [15]

$$\hat{b}_q = \frac{1}{\sqrt{N}}\hat{a}^+\hat{b}, \quad \hat{b}_q^+ = \frac{1}{\sqrt{N}}\hat{a}\hat{b}^+. \tag{11}$$

Gardiner's phonon operators obey the deformed commutation relation

$$\left[\hat{b}_q, \hat{b}_q^+\right] = 1 - \frac{2}{N}\hat{b}^+\hat{b} = 1 - 2\eta\hat{b}^+\hat{b}, \tag{12}$$

with $\eta = \frac{1}{N}$. When $\eta \to 0$ or $N \to \infty$, the standard (nondeformed) bosonic commutation relation is regained. In general the deformed operator $\hat{b}_q$ is related to nondeformed operator $\hat{b}$ through an operator valued function $f_1$ as

$$\hat{b}_q = \hat{b}f_1(\hat{b}_e^+\hat{b}_e;\eta). \tag{13}$$

In our particular case, we have

$$f_1(\hat{b}_e^+\hat{b}_e;\eta) = \sqrt{1 - \eta(\hat{b}_e^+\hat{b}_e - 1)}. \tag{14}$$

Here the deformation parameter is no longer phenomenological and is defined by the total number of atoms. For small deformation, the deformed boson operators $\hat{b}_q^+$ and $\hat{b}_q$ could be expressed in terms of the standard bosonic operators $\hat{b}^+$ and $\hat{b}$ ($[\hat{b},\hat{b}^+] = 1$) as

$$\hat{b}_q \approx \hat{b}[1-\frac{\eta}{2}(\hat{b}^+\hat{b}-1)] = \hat{b} - \frac{1}{2N}\hat{b}^+\hat{b}\hat{b},$$

$$\hat{b}_q^+ \approx [1-\frac{\eta}{2}(\hat{b}^+\hat{b}-1)]\hat{b}^+ = \hat{b}^+ - \frac{1}{2N}\hat{b}^+\hat{b}^+\hat{b}.$$
(15)

Now, we consider the effect of collisions between the atoms within the condensate as a special kind of $f$-deformation. For this purpose, we remind the basic about $f$-deformed oscillators in the following subsection.

## 2-1 $f$-Deformed oscillator algebra

The $q$-deformed oscillator algebra is determined by the following commutation relations for creation and annihilation operators $\hat{A}^+, \hat{A}$ and the number operator $\hat{N}$

$$[\hat{N},\hat{A}] = -\hat{A} \qquad [\hat{N},\hat{A}^+] = \hat{A}^+,$$
(16)

and the nonlinear relations

$$\hat{A}^+\hat{A} = [\hat{N}] \qquad \hat{A}\hat{A}^+ = [\hat{N}+1],$$
(17)

where the notation $[x]$ is defined as

$$[x] = \frac{q^x - q^{-x}}{q - q^{-1}},$$
(18)

and $q$ is the deformation parameter. A four-parameter generalized deformed algebra has been introduced in Ref. [29]

$$\hat{A}\hat{A}^+ - q^{\gamma}\hat{A}^+\hat{A} = q^{\alpha\hat{N}+\beta} \qquad [\hat{N},\hat{A}] = -\hat{A} \qquad [\hat{N},\hat{A}^+] = \hat{A}^+.$$
(19)

where $\alpha, \beta$ and $\gamma$ are real parameters. Man'ko and coworkers [25] introduced the $f$-deformed oscillator operators, defined as a nonlinear expansion of the standard harmonic oscillator operators $\hat{a}$ and $\hat{a}^+$,

$$\hat{A} = \hat{a}f(\hat{N}), \hat{A}^+ = f^*(\hat{N})\hat{a}^+, \hat{N} \equiv \hat{a}^+\hat{a}.$$
(20)

The function $f(\hat{N})$ is specific to each deformed algebra. This function determines the form of nonlinearities of the system under consideration and depends on four parameters $\alpha, \beta, \gamma$ and $q$. As an example, we consider the Hamiltonian for the free $f$-oscillator

$$\hat{H}(\hat{N}) = \frac{\hbar\omega_0}{2}(\hat{A}^+\hat{A} + \hat{A}\hat{A}^+) = \frac{\hbar\omega_0}{2}[\left|f(\hat{N}+1)\right|^2(\hat{N}+1) + \left|f(\hat{N})\right|^2\hat{N}].$$
(21)

Using the algebra (19) and deformation (20) we get

$$\left|f(\hat{N})\right|^2 = \begin{cases} \dfrac{q^\beta}{\hat{N}}\dfrac{q^{\alpha\hat{N}} - q^{\gamma\hat{N}}}{q^\alpha - q^\gamma}, & \alpha \neq \gamma \\ q^{\beta+\gamma(\hat{N}-1)}, & \alpha = \gamma \end{cases}.$$

(22)

By introducing new deformation parameters $q = e^\tau, \alpha = \upsilon + \mu, \gamma = \upsilon - \mu,$ we obtain

$$\left|f(\hat{N})\right|^2 = \frac{\sinh(\tau\mu\hat{N})}{\hat{N}\sinh(\tau\mu)}\exp\{\tau[\beta + \upsilon(\hat{N}-1)]\},$$
(23)

and the Hamiltonian of the free $f$-oscillator (21) can be written explicitly as

$$\hat{H} = \frac{\hbar\omega_0}{2} e^{\tau(\beta+\upsilon\hat{N})} \{\frac{\sinh(\tau\mu[\hat{N}+1])}{\sinh(\tau\mu)} + e^{-\tau\upsilon}\frac{\sinh(\tau\mu\hat{N})}{\sinh(\tau\mu)}\}. \tag{24}$$

We shall apply these results in the following subsection to illustrate the atomic collisions within the condensate as a specific kind of $f$-deformation.

## 2-2 An example: the collisions effect

As a particular physical example, we consider the atomic collisions within the condensate. The effective interaction Hamiltonian contains a nonlinear term proportional to $(\hat{a}^+\hat{a})^2$ [30, 31],

$$\hat{H}_I = \frac{\hbar\kappa}{2}(\hat{a}^+\hat{a})^2, \tag{25}$$

where the collision rate is denoted by $\kappa$. From kinetic theory, the collision rate is $\kappa \approx \rho\pi a^2 v_{rms}$, where $\rho$ is the density of the atoms, $a$ is the scattering length and $v_{rms}$ is the root-mean square-speed of the atoms. By expanding the Hamiltonian (21) and considering small values of $\upsilon$ and $\mu^2$ we obtain

$$\hat{H}(\hat{N}) = \frac{\hbar\omega_0}{2}[(2\hat{N}+1) + \frac{1}{6}\mu^2\hat{N} + (\frac{1}{2}\mu^2 + 2\upsilon)\hat{N}^2 + O(\upsilon^2, \upsilon\mu^2, \mu^4)] =$$
$$\frac{\hbar\omega_0}{2}[(2\hat{N}+1) + (\frac{2}{3}\mu^2 + 2\upsilon)\hat{N} + (\frac{1}{2}\mu^2 + 2\upsilon)\hat{N}(\hat{N}-1) + O(\upsilon^2, \upsilon\mu^2, \mu^4)], \tag{26}$$

where the interaction Hamiltonian reads as

$$\hat{H}_I(\hat{N}) \approx \frac{\hbar\omega_0}{2}[(\frac{1}{6}\mu^2\hat{N} + (\frac{1}{2}\mu^2 + 2\upsilon)\hat{N}^2]. \tag{27}$$

The Hamiltonian (27) reproduces the Hamiltonian (25) by setting $\mu^2 = 0$ and $\upsilon = \frac{\kappa}{2\omega_0}$. Thus, we see that the atomic collisions effect transforms the standard (nonlinear) harmonic oscillator model into an $f$-deformed one. Alternatively, we could set, *ab initio* in Eq. (21), $f(\hat{N}) = \sqrt{\kappa\hat{N}+(1-\kappa)}$ to obtain the Hamiltonian (25). Therefore, the parameters of the generalized deformed algebra are related to the rate of atomic collisions $\kappa$. As an interesting point we note that when $\mu^2 = -3\upsilon$ and $\upsilon = \frac{2k}{\omega_0}$ the Hamiltonian (26) reproduces the Kerr-like Hamiltonian, where the monochromatic field Hamiltonian contains, to lowest order, a nonlinear term proportional to $N(N-1)$

$$\hat{H}_{Kerr}(\hat{N}) = \frac{\hbar\omega_0}{2}(2\hat{N}+1) + \frac{k}{2}\hat{N}(\hat{N}-1). \tag{28}$$

Therefore one can infer that up to the first order approximation the nonlinearity of the models under consideration due to the atomic collisions within the condensate may be described as Kerr-type nonlinearity.

Subsequently, by considering the effect of collisions between the atoms within condensate, we can apply the extra deformation on the intrinsically deformed Gardiner's phonon operators for BEC by an operator-valued function $f_2(\hat{n}) = \sqrt{\kappa\hat{n}+(1-\kappa)}$ of the particle number operator $\hat{n}$. Here the nonlinearity is related to the collisions between the atoms within condensate. The operator valued-function $f_2(\hat{n})$ reduces to 1 as soon as $\kappa \to 0$. It means that the deformation increases with the collision rate $\kappa$.

The deformed Gardiner's phonon operators $\hat{B}_q$ and $\hat{B}_q^+$ are related to the operators $\hat{b}_q$ and $\hat{b}_q^+$ thorough the operator valued-function $f_2(\hat{n})$ as

$$\hat{B}_q = \hat{b}_q f_2(\hat{n}), \quad B_q^+ = f_2^+(\hat{n})\hat{b}_q^+, \qquad (29)$$
$$\hat{n} = \hat{N}_q = \hat{b}_q^+ \hat{b}_q.$$

Therefore the deformed version of the Hamiltonian (2) can be written as

$$\hat{H} = \hbar\overline{\omega}\hat{B}^+\hat{B} + \hbar\sqrt{N_c}[g(t)\hat{B}^+ + g^*(t)\hat{B}] + \sum_k \Omega_k \hat{c}_k^+ \hat{c}_k + \hbar\sqrt{N_c}\sum_k \xi(k)[\hat{b}^+\hat{c}_k + \hat{c}_k^+\hat{b}]. \qquad (30)$$

Note that, in the last term of r.h.s of Eq. (30), the nonlinear character of $\hat{B}_q$ has been neglected due to the weak-coupling assumption with the reservoir.

For small deformation, we obtain

$$\hat{B}_q = \hat{b}_q[1 - \frac{\kappa}{2}(1 - \hat{b}_q^+\hat{b}_q)], \quad \hat{B}_q^+ = [1 - \frac{\kappa}{2}(1 - \hat{b}_q^+\hat{b}_q)]\hat{b}_q^+. \qquad (31)$$

By keeping only the lowest order of $\eta = \frac{1}{N}$ for very large total number of atoms $N$ and by keeping only the first-order term of the collision rate $\kappa$ for very low temperature we get

$$\hat{B}_q \approx (\hat{b} - \frac{1}{2N}\hat{b}^+\hat{b}\hat{b})[1 - \frac{\kappa}{2}\{1 - (\hat{b}^+\hat{b} - \frac{1}{N}\hat{b}^+\hat{b}^+\hat{b}\hat{b})\}],$$
$$\hat{B}_q^+ \approx [1 - \frac{\kappa}{2}\{1 - (\hat{b}^+\hat{b} - \frac{1}{N}\hat{b}^+\hat{b}^+\hat{b}\hat{b})\}](\hat{b}^+ - \frac{1}{2N}\hat{b}^+\hat{b}^+\hat{b}). \qquad (32)$$

By assuming all the atoms are initially in the condensate phase we get $N = N_c$ and thus by using Eq. (32) the deformed Hamiltonian (30) can be expressed in terms of the nondeformed operators $\hat{b}$ and $\hat{b}^+$ as follows

$$\hat{H}_{eff} = \hbar\overline{\omega}\hat{b}^+\hat{b} + \hbar\overline{\omega}(\kappa - \eta)\hat{b}^+\hat{b}^+\hat{b}\hat{b} + \hbar\sqrt{N}[g(t)\hat{b}^+ + H.c.] +$$
$$\hbar\sqrt{N}[g(t)(\frac{\kappa - \eta}{2})\hat{b}^+\hat{b}^+\hat{b} + H.c.] + \sum_k \Omega_k \hat{c}_k^+ \hat{c}_k + \hbar\sqrt{N}\sum_k \xi(k)[\hat{b}^+\hat{c}_k + H.c.]. \qquad (33)$$

Now, we consider the Heisenberg equations of motion for the relevant operators with the Hamiltonian (33)

$$\dot{\hat{b}} = -\frac{i}{\hbar}[\hat{b}, \hat{H}_{eff}] = -i\overline{\omega}\hat{b}(t) - 2i\overline{\omega}(\kappa - \eta)\hat{b}^+(t)\hat{b}^2(t) -$$
$$ig(t)\sqrt{N} - ig(t)\sqrt{N}(\frac{\kappa - \eta}{2})(2\hat{b}^+(t)\hat{b}(t) + \hat{b}^2(t)) - i\sqrt{N}\sum_k \xi(k)\hat{c}_k(t), \qquad (34)$$

$$\dot{\hat{c}}_k = -\frac{i}{\hbar}[\hat{c}_k, \hat{H}_{eff}] = -i\Omega_k \hat{c}_k(t) - i\sqrt{N}\xi(k)\hat{b}(t). \qquad (35)$$

The equation of motion for the reservoir operator $\hat{c}_k(t)$ can be formally integrated to yield

$$\hat{c}_k(t) = \hat{c}_k(0)e^{-i\Omega_k t} - i\sqrt{N_c}\xi(k)\int_0^t dt' \hat{b}(t')e^{-i\Omega_k(t-t')}, \tag{36}$$

where, the first term describes the free evolution of the reservoir modes and the second term arises from their interaction with the harmonic oscillator. The reservoir operator $\hat{c}_k(t)$ can be removed by substituting the formal solution of $\hat{c}_k(t)$ into Eq. (34). By using the definition (3) we obtain

$$\dot{\tilde{b}}(t) = -i\Delta\tilde{b}(t) - 2i\overline{\omega}(\kappa-\eta)\tilde{b}^+\tilde{b}^2 - ig\sqrt{N} - ig\sqrt{N}\frac{(\kappa-\eta)}{2}(2\tilde{b}^+\tilde{b}+\tilde{b}^2)$$
$$+ f_{\tilde{b}}(t)\sqrt{N} - N\sum_k \xi^2(k)\int_0^t dt'\tilde{b}(t')e^{-i(\Omega_k-\Omega)(t-t')}, \tag{37}$$

$$\hat{f}_{\tilde{b}}(t) = -i\sum_k \xi(k)\hat{c}_k(0)e^{-i(\Omega_k-\Omega)t}. \tag{38}$$

Here, $\hat{f}_b(t)$ is a noise operator because it depends on the reservoir operator $\hat{c}_k(0)$. The fluctuations in the expectation values involving the harmonic oscillator operator will therefore depend on the evolution of the reservoir operators. The noise operator varies rapidly due to the presence of all the reservoir frequencies.

In view of the Weisskopf-Wigner approximation, the summation in Eq. (37) yields as $\delta(t-t')$ function and the integration can then be carried out. Therefore we obtain

$$N\sum_k \xi^2(k)\int_0^t dt'\tilde{b}(t')e^{-i(\Omega_k-\Omega)(t-t')} \cong \frac{1}{2}\varsigma\tilde{b}(t)N, \tag{39}$$

where, $\varsigma$ is the damping constant and defined as
$$\varsigma = 2\pi[\xi(\Omega)]^2 D(\Omega). \tag{40}$$

Here, the density of states $D(\Omega) = \frac{V\Omega^2}{\pi^2 c^3}$, with $V$ as the quantization volume, and $\xi(\Omega) \equiv \xi_{\frac{\Omega}{c}}$ is the coupling constant evaluated at $k = \frac{\Omega}{c}$. It is possible to make the substitutions $\frac{1}{2}\varsigma\tilde{b}(t)N_c = \Gamma\tilde{b}(t)$ and $\hat{f}_{\tilde{b}}(t)\sqrt{N_c} = \sqrt{2\Gamma}\hat{b}_{in}(t)$, so we can rewrite Eq. (37) in the following form

$$\dot{\tilde{b}}(t) = -i\Delta\tilde{b}(t) - 2i\overline{\omega}(\kappa-\eta)\tilde{b}^+\tilde{b}^2 - ig\sqrt{N} - ig\sqrt{N}\frac{(\kappa-\eta)}{2}(2\tilde{b}^+\tilde{b}+\tilde{b}^2)$$
$$+ \sqrt{2\Gamma}\hat{b}_{in}(t) - \Gamma\tilde{b}(t). \tag{41}$$

As $\kappa,\eta \to 0$ the linear equation (5) is recovered. In steady-state by using the semiclassical approximation, $\tilde{b}(t) = \beta + \delta\tilde{b}(t)$ we get

$$-i\Delta\beta - 2i\overline{\omega}(\kappa-\eta)\beta|\beta|^2 - i\sqrt{N}g - 2i\sqrt{N}g\frac{(\kappa-\eta)}{2}|\beta|^2 - i\sqrt{N}g\frac{(\kappa-\eta)}{2}\beta^2 - \Gamma\beta = 0 \tag{42}$$

By solving the above equation, we get the steady-state value of the field. The solution of the Eq. (42) reduces to Eq. (7) as soon as $\kappa,\eta \to 0$, as depicted in Figs 1 and 2.

By keeping only the first order term of the fluctuation $\delta\tilde{\hat{b}}(t)$ the dynamics of the small fluctuations is given by

$$\dot{\delta\tilde{\hat{b}}}(t) = A\delta\tilde{\hat{b}}(t) + B\delta\tilde{\hat{b}}^{+}(t) + \sqrt{2\Gamma}\hat{b}_{in}(t), \tag{43}$$

where

$$A = -i\Delta - \Gamma + i\sqrt{N}g(\eta - \kappa)(\beta + \beta^*) - 4i\overline{\omega}(\eta - \kappa)\beta^2,$$
$$B = i\sqrt{N}g(\eta - \kappa)\beta + 2i\overline{\omega}(\eta - \kappa)\beta^2. \tag{44}$$

Therefore, the Fourier component of the operator $\delta\tilde{\hat{b}}(t)$ reads

$$\delta\tilde{\hat{b}}(\omega) = \frac{1}{E(\omega)}\{[i\omega - A^*]\hat{b}_{in}(\omega) + B\hat{b}_{in}^{+}(\omega)\}, \tag{45}$$

where

$$E(\omega) = |A|^2 - |B|^2 - \omega^2 - i\omega(A + A^*). \tag{46}$$

Finally, by using Eqs. (6) and (45) the spectrum of light scattered from the $f$-deformed BEC under consideration is obtained as

$$S(\omega) = \int d\omega' \left\langle \delta\tilde{\hat{b}}^{+}(\omega)\delta\tilde{\hat{b}}(\omega') \right\rangle = \frac{|B|^2}{|E(\omega)|^2}. \tag{47}$$

When the limit $\kappa, \eta \to 0$ is taken, the result of Eq. (10), $S(\omega) = 0$, is recovered.

In Fig 3, we show the three-dimensional plot of the spectrum $S$ as a function of normalized frequency $\omega/\gamma$ and total number of atoms $N$ in the absence of atomic collisions ($\kappa = 0$). We see that the spectrum $S$ vanishes for $\eta \to 0 (N \to \infty)$. The spectrum $S$ shows a peak with decreasing the number of atoms, since the increasing in deformation parameter $\eta$ results in large nonlinearity, which may lead to observable effects on the spectrum of the scattered light. To observe the collisions effect, for a given number of atoms, we need to plot the spectrum $S$ as a function of normalized frequency $\omega/\gamma$ and the collision rate $\kappa$. This is shown in Fig 4. We see that the spectrum $S$ increases with the deformation parameter $\kappa$. The reason is due to the fact that considering the effect of collisions between the atoms within the BEC requires a deformation of the bosonic field [26]. Hence the presence of nonlinearity leads to a deviation from the usual predicted spectrum's shape. In the absence of atomic collisions the results of Ref. [13] is recovered.

## 3 Summary and conclusions

In summary, we have studied the spectrum of light scattered from an $f$-deformed BEC of a gas of two-level atoms in which the Gardiner's phonon operators are deformed by an operator-valued function $f(\hat{n})$, of the particle-number operator $\hat{n}$. By considering the effect of collisions between the atoms within the condensate, we have applied the extra deformation on the intrinsically deformed Gardiner's phonon operators for BEC. We have found that the presence of deformation parameters $\eta$ and $\kappa$ introduce nonlinearity, which may lead to observable effects on the spectrum of light scattered. Also, we have found that the deformation parameters $\eta$ and $\kappa$ play an important role in determining the spectrum of light scattered from the BEC. The scattering spectrum vanishes for $\kappa, \eta \to 0$. i. e., in the absence of deformations the usual predicted spectrum's shape is recovered. As pointed out before, the collision rate is $\kappa \approx \rho \pi a^2 v_{rms}$. Therefore we can adjust the value of the collision rate $\kappa$ by changing the density of the BEC. For example, lowering the temperature of the BEC increases the condensate density $\rho$, hence increases the collision rate $\kappa$, leading to increasing of deformation. The $f$-deformed BEC exhibits nonlinear characteristics and the nonlinearity leads to a deviation from the usual predicted spectrum's shape.

**References**

[1] M. H. Anderson, J. R. Ensher, M. R. Matthews, C.E. Wieman and E. A. Cornell, Science **269**, 198 (1995).

[2] C. C. Bradley, C. A. Saskett, J. J. Toellet and R. G. Hulet, Phys. Rev. Lett. **75**, 1687 (1995).

[3] K. B. Davis, M. O. Mewes, M. R. Anderson, N. J. van Durten, D. S. Durfee, D. M. Kurn and W. Ketterle, Phys. Rev. Lett. **75**, 3969 (1995).

[4] T. C. Killian, D. G. Fried, L. Willmann, D. Landhuis, S. C. Moss, T. J. Greytak and D. Kleppner, Phys. Rev. Lett. **81**, 3807 (1998).

[5] A. Robert, O. Sirjean, A. Browaeys, J. Poupard, S. Nowak, D. Boiron, C. I. Westbrook and A. Aspect, Science **292**, 461 (2001).

[6] J. Javanainen, Phys. Rev. Lett.**75**, 1927 (1995).

[7] R. Graham and D. F. Walls, Phys. Rev. Lett. **76**, 1774 (1996).

[8] B. Svistunov and G. Shlyapnikov, Zh. Eksp. Teor. Fiz. **97**, 821 (1990).

[9] H. D. Politzer, Phys. Rev. A. **43**, 6444 (1991).

[10] J. Javanainen, Phys. Rev. Lett.**72**, 2375 (1994).

[11] L. You, M. Lewnestein and J. Cooper, Phys. Rev. A **50**, R3565 (1994).

[12] Y. B. Band, Brazilian J. Phys. **27**, 147 (1996).

[13] S. Mancini and V. I. Man'ko, Phys. Lett. A **259**,66 (1999).

[14] N. N. Bogoliubov, J. Phys. (USSR) **11**, 23 (1947).

[15] C. W. Gardiner, Phys. Rev. A **56**, 1414 (1997).

[16] X. X. Liu, C. P. Sun, S. X. Yu and D. L. Zhou, Phys. Rev. A **63**, 023802 (2001).

[17] V. Pasquier, H. Saleur, Nucl. Phys. B **330**, 523 (1990).

[18] J. Wess and B. Zumino, Nucl. Phys. B **18**, 302 (1990).

[19] K. D. Sviratcheva, C. Bahri, A. I. Georgieva and J. P. Draayer, Phys. Rev. Lett. **93**, 152501 (2004).

[20] L. Alvarez-Gaume, C. Gomez and G. Sierra, Nucl. Phys. B **330**, 347 (1990).

[21] L. C. Biedenharn, J. Phys. A: Math. Gen. **22**, L873 (1989).

[22] A. J. Macfarlane, J. Phys. A: Math. Gen. **22**, 4581 (1989).

[23] M. Chaichian and P. Kulish, Phys. Lett. B **234**, 72 (1990).

[24] V. I. Man'ko, G. Marmo, S. Solimeno and F. Zaccaria, Int. J. Mod. Phys. A **8**, 3577 (1993).

[25] V. I. Man'ko, G. Marmo, S. Solimeno and F. Zaccaria, Phys.Lett. A **176**, 173 (1993).

[26] V. I. Man'ko, G. Marmo, F. Zaccaria and C. G. Sudarshan, Physica Scripta **55**, 523 (1997).

[27] M. H. Naderi, M. Soltanolkotabi and R. Roknizadeh, J. Phys. Soc. Japan , **73**, 2413 (2004).

[28] Z. Haghshenasfard, M. H. Naderi and M. Soltanolkotabi, eprint quant-ph/ 0801.2440.

[29] V. V. Borzov, E. V. Damaskinsky and S .B. Yegorov, J. Math. Sci. **100**, 2061 (2000).

[30] T. Wong, M. J. Collett and D. F. Walls, Phys. Rev. A **54**, R3718 (1996).

[31] C. W. Gardiner, A. S. Parkins and P. Zoller, Phys. Rev. A **46**, 4363 (1992).

**Figure captions**

Fig. 1. The quantity $\||\beta|-|\beta_0|\|$ as a function of the total number of atoms $N$ for $\kappa=0$ (no collision) and $\Delta=0$, $g=2.5\gamma, \overline{\omega}/\gamma=50$ and $\arg(\beta)=\arg(\beta_0)=\pi/2$.

Fig. 2. The quantity $\||\beta|-|\beta_0|\|$ as a function of the collision rate $\kappa$, when the total number of atoms $N=100$ and the values of other parameters are the same as those in Fig. 1.

Fig. 3. The spectrum $S(\omega)$ of light scattered from BEC as a function of the total number of atoms $N$ and the normalized frequency $\omega/\gamma$ in the case of no collision ($\kappa=0$). Other parameters are the same as those in Fig. 1.

Fig. 4. The spectrum $S(\omega)$ of light scattered from BEC as a function of the collision rate $\kappa$ and the normalized frequency $\omega/\gamma$ when the total number of atoms $N=100$. Other parameters are the same as those in Fig. 1.

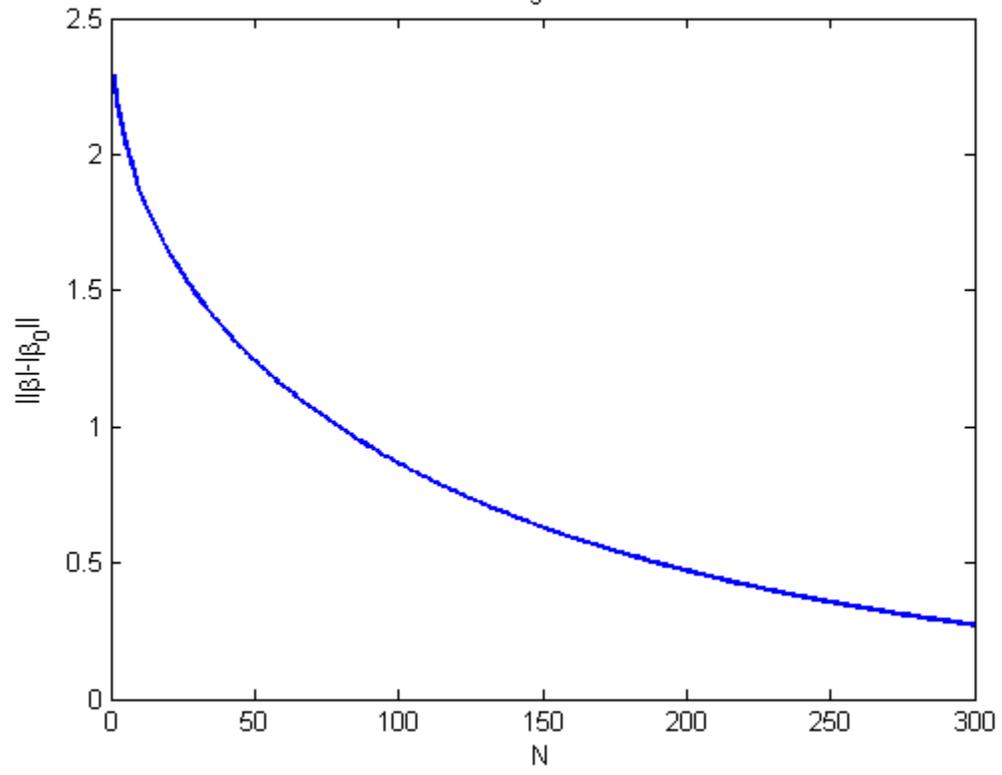

Fig. 1

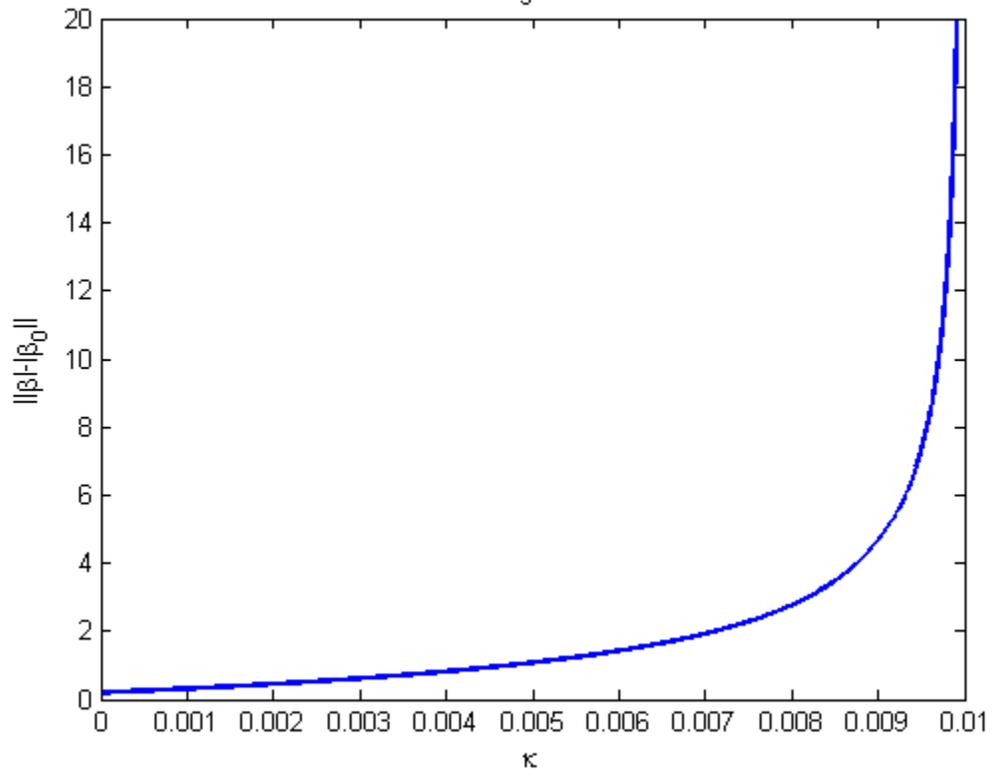

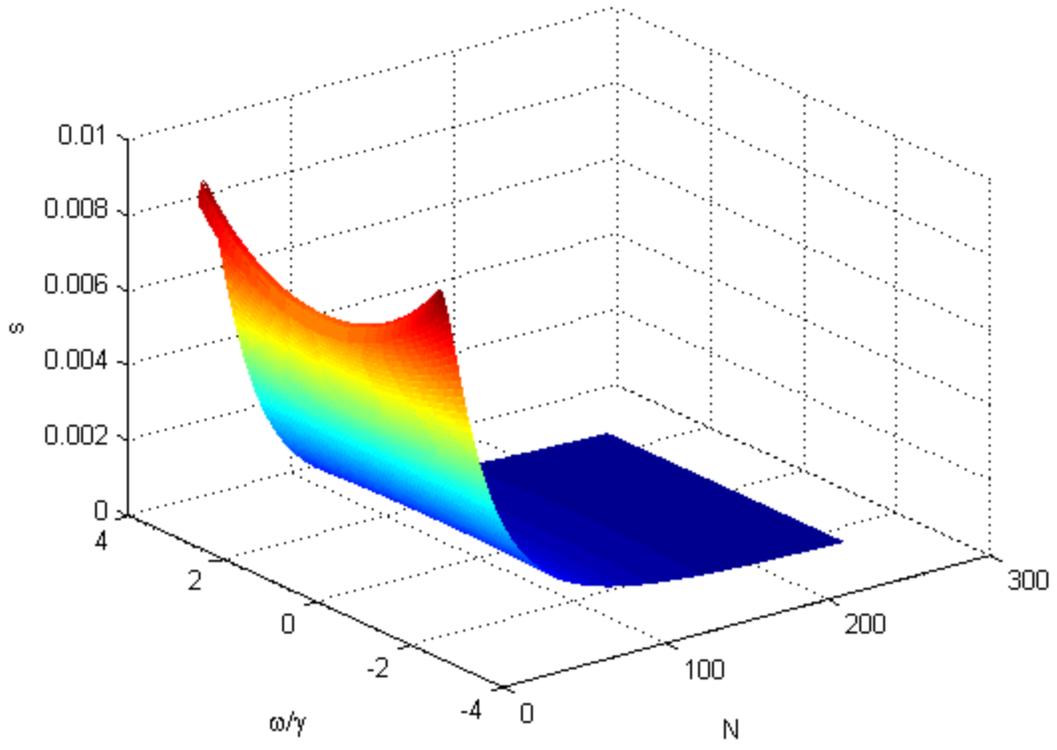

Fig. 3

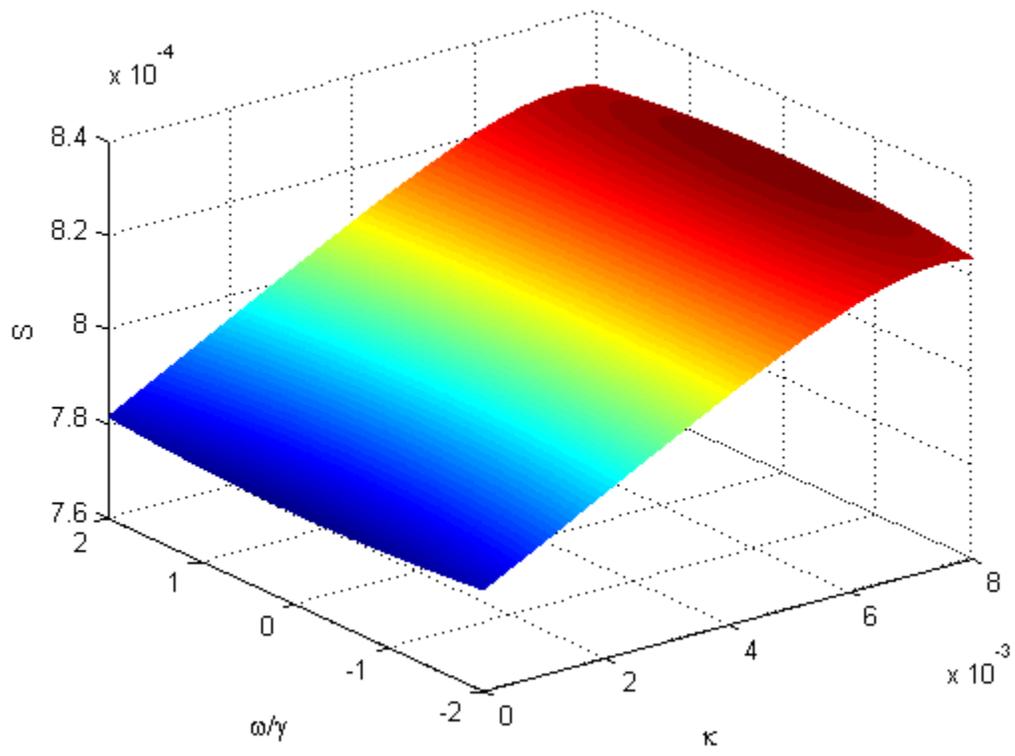

Fig. 4